\documentclass[pre,aps,preprint,showpacs,floatfix]{revtex4}
\usepackage{graphicx}
\usepackage{color}

\newcommand{\eq}[1]{Eq. ~(\ref{#1})}  
  
\newcommand{\be}[1]{\begin{equation}\label{#1}}  
\newcommand{\ee}{\end{equation}}  
\newcommand{\kick}{\langle\Delta t\rangle}

\begin{document}
\title{Dissociation and ionization of small molecules steered by external noise}
\author{Anatole Kenfack$^{(a)}$, Jan M. Rost$^{(a)}$, and Frank Gro\ss mann$^{(b)}$}

\date{\today}

\affiliation{(a) Max-Planck-Institut f\"ur Physik Komplexer Systeme, N\"othnitzer Strasse 38, D-01187 Dresden, Germany\\
(b) Insitute for Theoretical Physics, Technical University of Dresden, 01062 Dresden}
\begin{abstract}
We show that ionization and dissociation can be influenced 
separately in a molecule with appropriate external noise.
Specifically we investigate the hydrogen molecular ion under a 
stochastic force  quantum mechanically beyond the Born-Oppenheimer 
approximation.  We find that up to 30$\%$ of
dissociation without ionization can be achieved by suitably tuning the
forcing parameters.
\end{abstract}
\pacs{33. 80. Gj, 34. 10. x, 03. 65. Sq}

\maketitle

\section{Introduction}

The anharmonicity of molecular vibrations makes the dissociation of a
molecule by irradiation of laser light a relatively difficult task
\cite{bloembergen}.  Consequently, high intensity is required for
dissociation, for instance, $I>10^{14}$W/cm$^2$ for HF in
monochromatic fields \cite{che90}.  In half cycle pulse fields the
situation is improving due to the broad frequency content of this
radiation.  Even in that case, however, intensities
$I>10^{13}$W/cm$^2$ are required to achieve on the order of 15$\%$
dissociation probability \cite{lin1}.  At such high intensities, the
ionization process dominates and masks vibrational excitation and
dissociation.  Chelkowsky et al.\ \cite{cheal90} suggested that the
dissociation threshold of a diatomic molecule can be lowered by two
orders of magnitude using a down-chirped laser, tracing the decrease
in spacing of the excited  energies levels, and hence dissociation
without ionization should be possible.  In a similar spirit,
circularly chirped pulses have been used by Kim et al.\ \cite{kim} for
the dissociation of diatomic molecules.  They found that the threshold
laser intensity is substantially reduced, which may allow to achieve
dissociation without ionization.

Recently, 
dissociation of diatomic molecules under a stochastic force was investigated
representing the diatom as a Morse oscillator \cite{jcp}. A 
phenomenon, akin to  stochastic resonance has also been identified in
the interaction of an atom with a strong laser field and additional 
noise \cite{kamal}. 

In the present study, the external force is a sequence of pulses
(kicks) at random times, each kick carrying an independent weight~
\cite{jcp,masoliver}.  Both, the weights $\Gamma_{i}$ and the
intervals between kicks $\Delta t_{i}$, follow a Poisson distribution
with averages $\gamma$ and $\langle \Delta t\rangle$, respectively.
This type of force, similar to white shot noise, has also been used to
model the passage of ions through carbon foils \cite{burgdorfer}.
Through its inherent time-structure characterized by $\langle \Delta
t\rangle$, it can couple selectively to bound
degrees of freedom with comparable time scales of the noise.  Given 
the difficult dissociation without ionization as described above, it 
is natural to ask if  stochastic forcing may allow
one to separate effectively ionization from dissociation.

In order to answer this question we have to include both, nuclear as
well as electronic degrees of freedom in our description.  The
simplest molecule displaying ionization and dissociation is the
hydrogen molecular ion H$_2^+$ in an external field.  Using a
two-level approximation for the electron \cite{lin2} or a soft-core
potential for the Coulomb interaction \cite{volal98,feth03}, this
system has been investigated previously under strong laser fields.

As in \cite{CZAB95} we refrain from these approximations and investigate
this molecule under the influence of a stochastic field with the only
restriction that the random force is directed along the molecular
axis.  This leads to cylindrical symmetry of the problem and therefore
3 degrees of freedom (2 for the electron and one for the nuclei) have
to be taken into account explicitly.  On the same level, the
interaction of H$_2^+$ with laser light has been investigated recently
and a sensitive dependence of the dissociation on the carrier envelope
phase has been found \cite{roudal04}.  Yet, the maximal dissociation
probability reported is below 20 $\%$.

We begin the main text of this paper in section 2 with a description
of the noise source and a brief explanation, how quantum 
evolution under a stochastic force is solved for in time. 
In section 3 we  discuss the influence
of the same stochastic force on an atom (electronic degrees of
freedom) and a molecule (vibrational degrees of freedom) separately,
since the effect of coupling strong noise to quantum systems is not
widely known.  Then in section 4 we describe how we handle H$_2^+$
numerically, in particular how we extract ionization and dissocication
from the numerically obtained time-dependent wavefunction.  In section
5, results for the dissociation and the ionization probability of
$H_2^+$ will be presented and discussed, while section 6 concludes
the paper.

\section{Description of the noise source}
The stochastic force $F(t)$ we consider here is given by \cite{masoliver,haenggi} 
\begin{eqnarray}  
F(t)=\sum_{i=1}^{N_t}\gamma_{i}\delta(t-t_{i}),\label{force}  
\end{eqnarray}  
and stands for a series of random impulses of strength $\gamma_{i}$ at 
times $t_{i}$, i.\ e.,  $F(t)$ is a kind of white shot noise  \cite{broeck} 
responsible for multiple $\delta-$kicks undergone by the molecule, 
where $N_t$ is the number of kicks up to time $t$ controlled by the 
Poisson counting process $N_t$.  It is characterized by the average kicking 
interval $\langle \Delta t\rangle \equiv \lambda^{-1}$ about which the 
actual interval $\Delta t_i=t_i-t_{i-1}$ are exponentially distributed, 
similarly as the actual kicking strengths $\gamma_i$ about  their 
mean $\gamma$,  
\be{deltat}  
P(\Delta t_i) = \lambda\exp(-\lambda\Delta t_i),\,\,P(\gamma_i) = \gamma^{-1}\exp(-\gamma_i/\gamma)\,.   
\ee  
  
In analogy to periodically applied half cycle pulses \cite{frey}, we
restrict our analysis to positive $\gamma_{i}$ and assume that
$\gamma_{i}$ and $t_{i}$ are mutually uncorrelated random variables
generated by the distributions functions of \eq{deltat}.  The
determination of $F(t)$ reduces to the construction of a stochastic
sequence $(\gamma_{i},t_{i})$ which can be done assuming that the
random times $t_{i}$ form a Poisson sequence of points leading to a
delta correlated process \cite{masoliver}.  It is easy to show
\cite{haenggi} that the stochastic force constructed has the
properties \begin{eqnarray} \langle F(t) \rangle&=&\gamma\lambda
\nonumber\\
    \langle 
    F(t)F(s)\rangle&=&2\gamma^2\lambda\delta(t-s)+\gamma^2\lambda^2\,,  \label{averageforce} 
\end{eqnarray} 
where $\langle \rangle$ is understood as an average over a sufficient 
large number  of deterministic realizations $j = 1,\ldots N$ of $F(t)$ in terms of 
specified stochastic sequences $(\gamma_{i}^{(j)},t_{i}^{(j)})$.
The corresponding power spectrum, i. e., the Fourier transform 
of $\langle F(t)F(s)\rangle$, is given by  
\begin{eqnarray} S(\omega) = 4\frac{\gamma^2\lambda}{\sqrt{2\pi}}+\gamma^2\lambda^2\sqrt{2\pi}\delta(\omega).  
\label{powerspectrum} 
\end{eqnarray} 
These properties reveal  the difference between the present stochastic 
force (white shot noise) and a pure white noise which is delta-correlated 
with {\it zero mean}. 

Note that $\langle \Delta t\rangle$ and $\gamma$ are the two relevant 
parameters characterizing the present noise source.

The determination of the time evolution of a system with a
deterministic Hamiltonian $H_{0}$ and an additional stochastic driving
$H = H_{0}+ xF(t)$ is straight forward by solving first the
conventional deterministic time-dependent Schr\"odinger equation for
each realization $F^{(j)}$ of the stochastic force, forming from it
the desired observable $\cal O^{(j)}$ and finally averaging over the
realizations, ${\cal O} = N^{-1}\sum_{j=1}^{N}{\cal O}^{(j)}$.

\section{Ionization and Dissociation under Stochastic Forces}

\begin{figure}
\begin{center}
\includegraphics[width=0.6\columnwidth,clip]{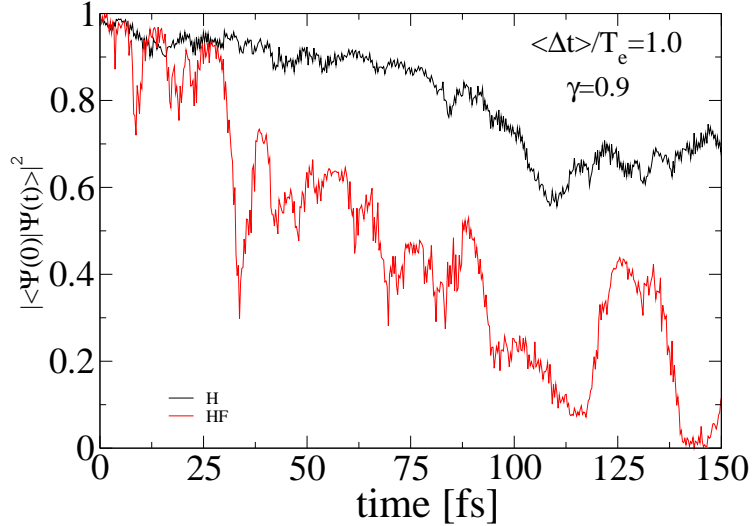}
\caption{\label{fig1}The population of the (electronic) ground state
of the hydrogen H atom versus the population of the vibronic ground
state of a Morse oscillator, describing the HF molecule as function of
time for the same amount of noise.  For HF model,
$\mu_0=3.54076$~a.u., $\alpha=1.1741$~a.u.,
$D=0.225$~a.u.\cite{goggin}.}
\end{center}
\end{figure}

To demonstrate that coupling white shot noise to a bound quantum
system is sensitive to the time scale of bound motion, we briefly
describe how an atom and a diatomic molecule responds to noise.  To
this end we show in figure \ref{fig1} how the electronic ground state
of a one dimensional soft-core model of the H-atom and the vibrational
ground state of a Morse oscillator describing a HF molecule are
depopulated under stochastic forcing.  The potential in the atom case
is given in atomic units by
\begin{eqnarray}
V(x,t)=V_C+xF_{\rm a}(t)=-\frac{1}{\sqrt{x^2+a}}+xF_{\rm a}(t),
\end{eqnarray}
with $a=2$ such that the ground state energy corresponds to that of
the 3d hydrogen atom.  In the molecular case it is given by
\begin{eqnarray}
V(x)=V_M+\mu_0 x F_{\rm m}(t)=D(1-\exp\{-\alpha x\})^2+\mu_0 x F_{\rm m}(t)
\end{eqnarray}
with the dissociation energy $D$, the length scale $\alpha$, and the
dipole gradient $\mu_0$.  Note that the eigenstates and the
eigenenergies of this Morse potential model are exactly known~\cite{ter46}.  We
expose both systems to the same stochastic external perturbation, as
described above, by adjusting the molecular force such that $\mu_0
F_{\rm m}(t)=F_{\rm a}(t)$.  The average time between kicks is chosen
to equal the electronic period in the hydrogen atom, which in atomic
units is given by $T_{\rm e}=2\pi$.

Numerically, we find a very promising indication that the electron in
the atom case does react much less to the force than the binding of
the molecule.  This is shown in figure \ref{fig1}, where the
auto-correlation function of the ground state electron wavefunction
for the atomic, as well as of the vibrational ground state in the
molecular case is shown.  While the atom is still dominantly in its
ground state, the ground state of the molecule has already been
depopulated by more than 50\%.

\begin{figure}
\begin{center}
\includegraphics[width=0.6\columnwidth,clip]{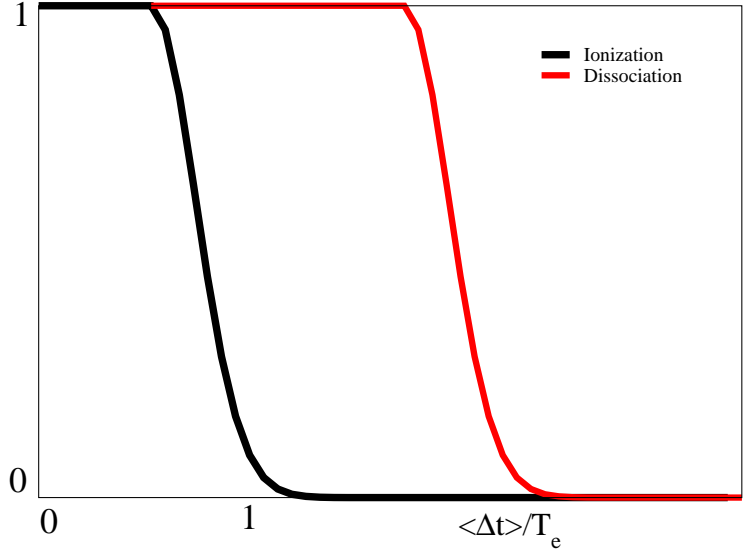}
\caption{\label{fig2}
Schematic ionization and dissociation (including fragmentation)
probabilities as function of the kick spacing $\langle \Delta
t\rangle/T_e$ at a given finite time and for a given kicking strength
$\gamma$.}
\end{center}
\end{figure}

In general, we expect for a system $S$ with a characteristic 
period $T_{S}^{c}$ of bound motion that noise will couple energy into 
the system if its kick spacing is smaller than the bond period, 
$\kick< T_{S}^{c}$. In the opposite case,  $\kick\gg T_{S}^{c}$, the 
bound motion will be insensitive to the noise. Since the molecular 
vibrational period is much larger than the electronic period in an 
atom,
$T^{c}_{\mathrm{vib}}\gg T^{c}_{\mathrm{e}}$, there is a 
window in $\kick$, where dissociation without substantial ionization 
should be possible (see the sketch, fig.~\ref{fig2}).

We double check this conjecture in the following with the hydrogen
molecular ion $H_2^+$ in an external field which is the simplest
molecule displaying both,  ionization and dissociation.  Since 
this problem is sensitive to different timescales, we avoid the
Born Oppenheimer approximation which could be an (artifical) source
of physical effects to be predicted.

\section{Molecular Ionization and Dissociation Dynamics in 
$H_{2}^{+}$ under external driving}  

We solve the time-dependent Schr\"odinger equation for the hydrogen
molecular ion in a linearly polarized laser field neglecting the center of
mass and the rotational motion \cite{CZAB95}.  Adding a dipole coupled stochastic force
$F(t)$ the time-dependent stochastic Schr\"odinger equation is given
by (in atomic units)
\begin{eqnarray}
i\frac{\partial }{\partial t}\Psi(z,\rho,R,t)=\left[
K_{\mathrm{vib}}+K_{\mathrm{e}}+V+z\kappa F(t)\right]\Psi(z,\rho,R,t)\label{sseq}
\end{eqnarray}
with the kinetic energy 
$K_{\mathrm{vib}}=-M^{-1}_{p}\partial^{2}/\partial R^{2}$ of the 
protons, the electronic kinetic energy
\begin{equation}
 K_{e} =   -\frac{\beta}{2}\frac{\partial^2}{\partial z^2}-\beta \frac{1}{2}\frac{\partial^2 }{\partial \rho^2}-\frac{1}{2\rho}
\frac{\partial}{\partial \rho}
\label{kine}
\end{equation}
and the potential energy
\begin{equation}
    V(\rho,z,R)=-[\rho^2+(z-R/2)^2]^{-1/2}-[\rho^2+(z+R/2)^2]^{-1/2}+1/R\,,
    \end{equation}
where $\beta=1/2+1/(4M_p)$, $\kappa=1+1/(2M_p+1)$, $M_p$ is 
 the proton mass in units of the electron mass, 
 $R$ is the internuclear distance and $z$ and $\rho$
are cylindrical coordinates of the electron.

Due to the cylindrical symmetry of the stochastically driven system an
expansion of the wavefunction in a Bessel-Fourier series in $\rho$ is
performed.  The singularities of the Coulomb potential is retained,
without  any softening of the potential.  The
time-dependent stochastic Schr\"odinger equation (\ref{sseq}) can
then be solved by the standard split operator FFT technique
\cite{feit}.  We start from the ground state as initial state $\Psi(\rho,z,R,t=0)$,
represented very accurately by  the
product of the vibrational ground state $\phi_0(R)$ and the
$1s\sigma_g$ electronic wavefunction of $H_2^+$.

\begin{figure}
\begin{center}
\includegraphics[width=0.6\columnwidth]{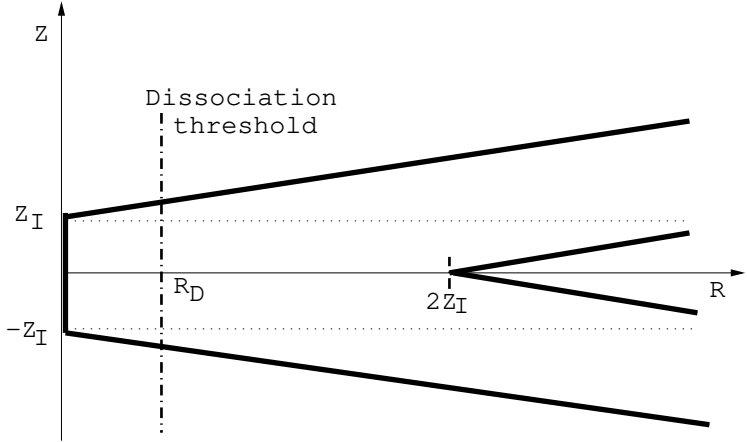}
\caption{The region inside the thick solid lines indicates the support
of the non-ionized part of the total wavefunction. The vertical
dot-dashed line defines the onset of dissociation.}
\label{fig3}
\end{center}
\end{figure}

The observables we are interested in are
the ionization probability  
\begin{eqnarray}
P_{\rm I}(t)=1-\int_{0}^{R_{\rm max}}f_1(R,t) dR
\end{eqnarray}  
and the dissociation probability which is defined 
{\it without} ionization according to
\begin{eqnarray}
P_{\rm D}(t)=\int_{R_{\rm D}}^{R_{\rm max}}f_1(R,t) dR\,.
\end{eqnarray} 
Here, $f_1$ is obtained by integrating $|\Psi(\rho,z,R,t)|^{2}$ over
electronic coordinates inside a cylinder of radius $\rho_{0}=8$ a.u.
and height 2$\times z_I+R$ as sketched in figure \ref{fig3}.  The
integration area would split into two cylinders if $R>2z_I$.  The
splitting, however, is never reached in the present calculation due to
the smallness of our $R$-grid.  For $z>z_I+R/2$, the system is
considered to be ionized since the electron is sufficiently distant
 to each  of the nuclei.  The nuclear separation defining the
onset of dissociation into the H+H$^+$ channel is taken as $R_D=9.5$
a.u. and \cite{CZAB95}.  In this configuration, $f_1$ is explicitely
given  by

\begin{eqnarray}
f_1(R,t)=2\pi\int_{-(z_I+R/2)}^{(z_I+R/2)}dz\int_{0}^{\rho_{0}} d\rho\rho |\psi(R,z,\rho,t)|^2\nonumber
\end{eqnarray}

In our calculations, the electronic grid for $z$ contains 1024 points 
and  extends from
$-50$~a.u. to $50$~a.u.. 
A quadratic imaginary potential has been
used to avoid unphysical reflections at the grid boundaries.  The
onset of ionization was defined to be at $z_I=32$ a.u..  The nuclear
grid consists of 256 points, extending from $R_{\rm min}=0.38$~a.u. to
$R_{\rm max}=24$~a.u. and 16 basis functions were used in the
Bessel-Fourier expansion for $\rho$.

\section{Numerical Results}

 Dissociation and ionization probabilities of $H_2^+$ as described in
 the previous section have been obtained for different parameters of
 the stochastic forcing.  Averaging over around 20 realizations of the
 noise was enough to converge the results for all the cases presented
 below.  The maximum propagation times for the wave function were
 always well below 500 fs around which which rotational motion of the
 molecule (not included in the present approach) would come into play.
\begin{figure}
\begin{center}
\includegraphics[width=0.6\columnwidth,clip]{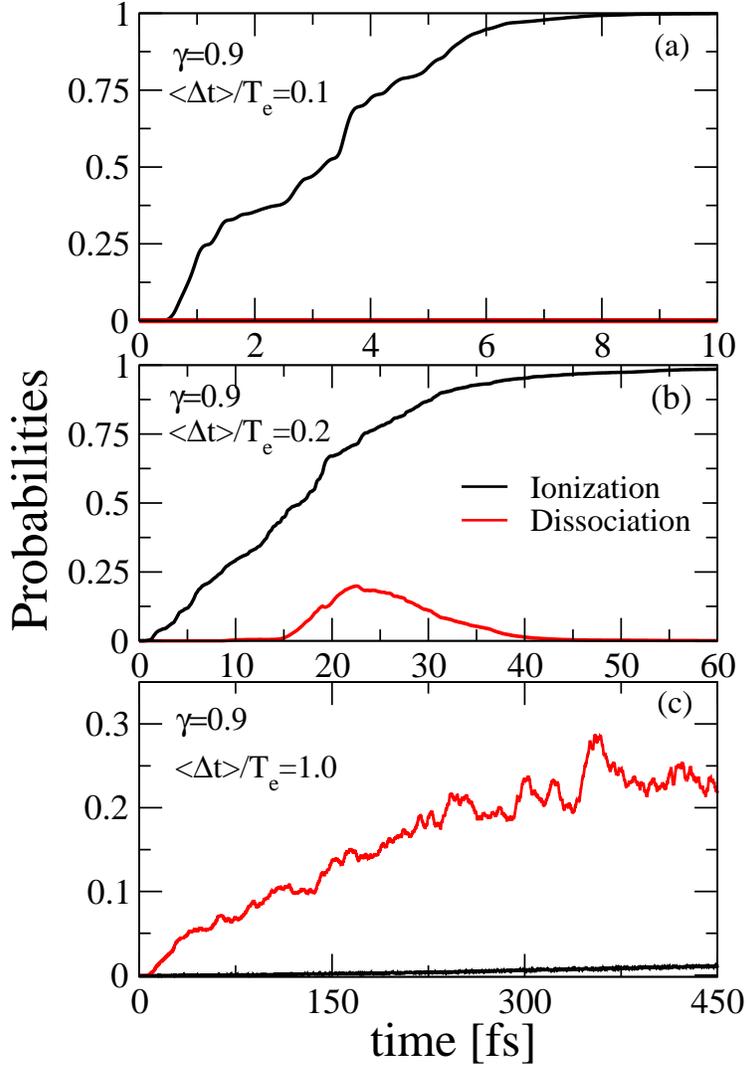}
\caption{Ionization and dissociation probabilities
as a function of time for different average kick spacings., with $\langle \Delta t\rangle/T_e$ as indicated in
panels and for a constant kicking strength $\gamma=0.9$~a.u.}
\label{fig4}
\end{center}
\end{figure}  

In figure \ref{fig4} we present  results for a fixed strength 
 $\gamma=0.9$~a.u. of the noise.
The average spacing of the kicks was varied to cover the  
switching of the dynamics from pure ionization 
($\langle\Delta t\rangle/T_e\ll 1$, figure \ref{fig4}a)
to pure dissociation  ($\langle\Delta t\rangle/T_e\approx 1$, figure 
\ref{fig4}c).
While  the ionization probability reaches unity the dissociation
probability  has a maximum of 25 $\%$ as can be seen in Figure \ref{fig4}c.

\begin{figure}
\begin{center}
\includegraphics[width=0.6\columnwidth,clip]{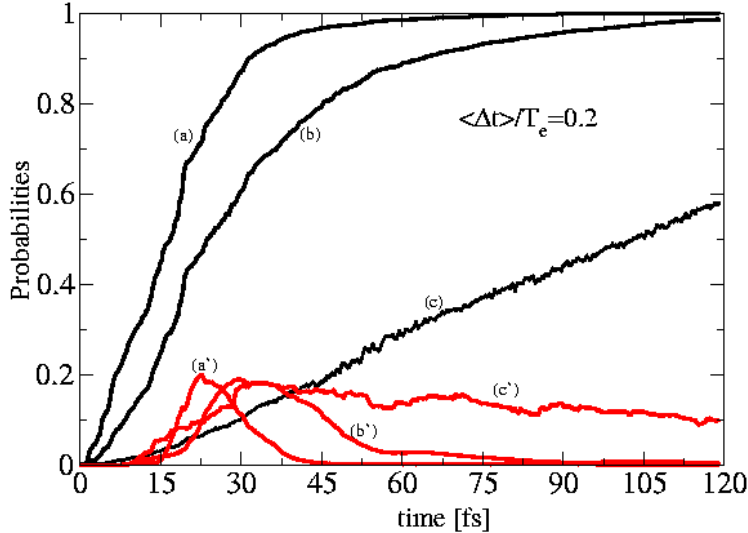}
\vspace{0.5cm}
\caption{Ionization (black) and dissociation (red)
probabilities as a function of time for different 
noise strength and for fixed spacing $\langle \Delta t\rangle/T_e=0.2$: 
ionization (a) and dissociation (a'): $\gamma=0.9$~a.u.
ionization (b) and dissociation (b'): $\gamma=0.75$~a.u.
ionization (c) and dissociation (c'): $\gamma=0.5$~a.u.}
\label{fig5}
\end{center}
\end{figure}

In Figure \ref{fig5} the spacing between the kicks is fixed at the
intermediate value of $\langle\Delta t\rangle/T_e=0.2$, where both,
dissociation and ionization, coexist for a certain amount of time.
From a) to c) we reduce the strength of the external forcing.
Reducing $\gamma$ reduces the ionization rate and correspondingly
shifts the peak in the dissociation without ionization to later times.

\begin{figure}
\begin{center}
\includegraphics[width=0.6\columnwidth,clip]{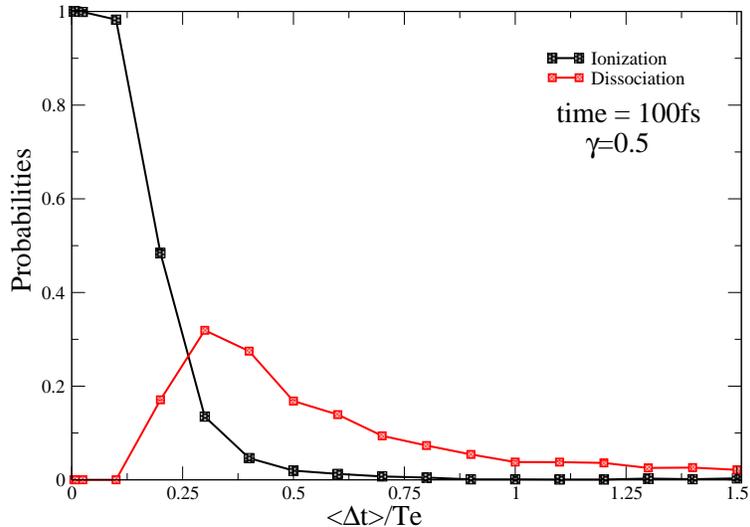}
\caption{Ionization and dissociation probabilities as a function of 
kick spacings at a given finite time for a kicking strength of
$\gamma=0.5$~a.u..}
\label{fig6}
\end{center}
\end{figure}

   What are close to optimal parameters for dissociation without
   ionization for the stochastically driven hydrogen molecular ion?
   This question is answered in figure \ref{fig6}.  Here, at fixed
   final time of 100 fs, the probabilities as a function of the
   average kick intervals are displayed.  Beyond $\langle\Delta
   t\rangle/T_e\approx 1$ the ionization quickly goes to zero.
   However, also the dissociation approaches zero, as the system does
   not respond to rare kicks.  An intermediate value of $\langle
   \Delta t\rangle/T_e\approx 0.5$ turns out to maximize dissociation
   without fragmentation of the whole system.  The maximal probability
   is close to 30 $\%$ and therefore higher than any comparable value
   we have found in the literature.  We did many additional
   calculations for different final times, as well as for different
   field strengths and found results qualitatively similar to those
   presented in Figure \ref{fig6}. 
   
   These results confirm the expectations from figure \ref{fig2} with 
   the important difference, that for the real molecule, the onset of 
   ionization limits the maximum dissociation which can be achieved. 
   The actual value of the maximum, of course, can only be calculated 
   as we have done (see figure \ref{fig6}).

\section{Discussion and Conclusions}

Achieving efficient dissociation in small molecular systems without 
ionization has been a
longstanding goal. Here, we have explored an alternative to the
standard tools, such as chirped laser pulses: 
the application of stochastic driving in the form of white shot noise 
with a characteristic internal timescale. 

For the hydrogen molecular ion we have shown that dissociation and
ionization can be well separated by a suitable choice of the noise
parameters.  A major role being played by the internal timescale of the noise,
the average interval between  kicks.  At kicking intervals equal to
the electronic ground state period of the hydrogen atom the driven
system switches between being ionized to dissociating already for
moderate field strengths below one atomic unit.

Stochastic driving with the specified properties can be realized with 
chaotic light, where a short laser pulse with sufficient band width 
is shaped with random phases and amplitudes of each frequency 
component \cite{kamal}.

\section*{Acknowledgment}
Fruitful discussions with Kamal P. Singh and Manfred Lein are 
gratefully acknowledged. A. K. would like to acknowledge the 
financial support by the Alexander von Humboldt-Stiftung as well as the 
Max-Planck-Gesellschaft through a Reimar L\"ust grant (2005).

\end{document}